\documentclass[preprint,12pt]{elsarticle}

\usepackage{matlab-prettifier}
\usepackage{lineno}
\usepackage{graphicx}
\usepackage{svg}
\usepackage{amsmath}
\usepackage{xfrac}
\usepackage{parskip}
\usepackage[backref=page]{hyperref}
\usepackage{adjustbox}
\usepackage{setspace}
\hypersetup{ 
backref=true, 
bookmarks=true, 
unicode=false, 
pdftoolbar=true, 
pdfmenubar=true, 
pdffitwindow=false, 
pdfstartview={FitH}, 
pdftitle={}, 
pdfauthor={}, 
pdfsubject={}, 
pdfkeywords={}{}, 
pdfnewwindow=true, 
colorlinks=true, 
linkcolor=BlueViolet, 
citecolor=maroon, 
urlcolor=blue,
}
\usepackage{comment}
\usepackage{subfig}
\usepackage{caption}
\usepackage{longtable}
\usepackage{amssymb}
\usepackage{gensymb}
\usepackage{siunitx}

\usepackage{natbib}
\biboptions{sort&compress}
\setcitestyle{square}

\begin{document}
\begin{frontmatter}

\title{An approach to extend Cross-Impact Balance method in multiple timespans
\footnote{The submission of this article aims to save an original archive as soon as possible in order to prevent any types of further infringement from a previous instructor. Special thanks to the student collaborators who worked together during the previous project and also suffered similar hardships. Special thanks also to Dr. Wolfgang Weimer-Jehle for answering my questions patiently and thoroughly during the email communication.}}

\author{Chonghao Zhao\corref{cor1}}

\cortext[cor1]{School of Mathematical Science, Beijing Normal University. No.19, Xinjiekouwai St, Haidian District, Beijing, 100875, P.R.China}

\ead{chzhao@mail.bnu.edu.cn}

\begin{abstract}
Cross-Impact Balance Analysis (CIB) is a widely used method to build scenarios and help researchers to formulate policies in different fields, such as management sciences and social sciences. During the development of the CIB method over the years, some derivative methods were developed to expand its application scope, including a method called dynamic CIB. However, the workflow of dynamic CIB is relatively complex. In this article, we provide another approach to extend CIB in multiple timespans based on the concept ‘scenario weight’ and simplify the workflow to bring convenience to the policy makers.

\end{abstract}

\end{frontmatter}

\newpage

\section{Introduction}
Cross-Impact Balance Analysis (CIB) is a qualitative-quantitative combined method to construct scenarios, which first introduced by Weimer-Jehle in 2006\cite{weimer2006}. The aim of CIB method is in one hand to weaken the subjectivity of expert judgements during the workflow of scenarios construction, on the other hand to combine qualitative and quantitative factors into a unified model to expand the application scope. The greatest advantage of the original CIB method lies in its straightforward and easy-to-understand operational process. The algorithm for finding scenarios also does not require users to have a high level of mathematical knowledge to comprehend\cite{weimer2006,weimer2009,weimer2023}.

The original CIB method mainly have 4 steps\cite{weimer2006}:
\begin{enumerate}
    \item Specifying the system we want to analyze and considering the decisive factors in this system. All these factors will be referred to as descriptors in the remaining part of this article.
    \item Specifying the possible states of the descriptors we chose. In the same descriptor, the states can be either all qualitative or quantitative.
    \item Specifying the interactions between the states and finally form a matrix called Cross-Impact Matrix (CIM). The interactions can be acquired by literature review or expert consultation, then quantified as integers from -3 to +3 (in most studies).
    \item Using consistency algorithm to check all possible scenarios in the system we constructed, all of the scenarios which passed the check are those we interested in and can be analyzed in further discussion.
\end{enumerate}
During the development of the CIB method over the years, some derivative methods were developed to expand its application scope. Linked CIB\cite{schweizer2016} and multi-level CIB\cite{vogele2017} both aim to deal with larger set of descriptors, split descriptor sets into several parts and apply consistency algorithm separately, then combine the scenarios of parts into the scenarios of the whole system. These two approaches can greatly shorten the time complexity of algorithm verification if different descriptors in independent levels do not have interactions. 

There are also some other approaches that focus on the timespan. Vögele (2019) raised an approach to expand CIB method in timespans, called dynamic CIB\cite{vogele2019}, which give some descriptors a threshold or cyclic behavior, then construct CIM in separate time periods if the descriptor reached a fixed state by the end of each period. Broska (2022) applied dynamic CIB and gave a detailed description in the research\cite{broska2022}. Schweizer and Jamieson-Lane (2023) combined CIB method with Markov process\cite{schweizer2023}, using graph and Markov chain to visualize the succession in CIB method, which can depict a chain of how to achieve a consistent scenario under either classic or stochastic succession rules.

However, the two approaches above have some problems in practical using. The workflow of dynamic CIB is complex and needs to construct CIM several times because whether a descriptor could touch the threshold or not cannot be predicted before the checking of consistency algorithm. CIB with Markov chain provided a way to present the route to a consistent scenario, however, it did not expand the timespans to the future. Thus, we need an approach to extend CIB method into the future with a concise workflow.

In this article, we suppose readers have basic knowledge of CIB method and all the terminologies will follow Weimer-Jehle (2009)\cite{weimer2009} in the remain parts. In section 2, we will extend the CIB in multiple timespans in another way different with dynamic CIB. In section 3, we will discuss the details of the method in section 2 and analyze its limitations. In appendix, we will formalize the method of scenarios aggregation in multi-level CIB for the purpose in the footnote in the first page of this article.

\section{Deal with CIB in multiple timespans}
For the dynamic CIB, we start from a set of static-threshold hybrid descriptor set and an original CIM for the specified first period of timespan, then use consistency algorithm to solve the CIM. If the states of the consistent scenarios of the first CIM reached the threshold, the researchers should construct another CIM for the second period. After repeating such process, we can construct a tree-like structure for the system we investigate. For the consistent scenarios which we interested in at the end of the tree (called leaves in graph theory), we can find a unique path to the original scenario. Those paths from ‘root’ to ‘leaves’ give researchers a possible evolution mode to the consistent scenarios, which can also give them a direction on policy making\cite{vogele2019}. However, the workflow of dynamic CIB is complex, as whether the threshold can be reached or not is not predictable due to the interaction between the states, which may increase the workload for both researchers and the experts in the workshop.

To decrease the complexity of the workflow, our aim is to find a modified workflow that can construct CIM at the same time but still can get a scenario chain at the end.
\begin{enumerate}
    \item First, we also start from the choosing of descriptors and corresponding states, then specify our research periods and split it into several timespans.
    \item Second, after the framework is determined, we construct CIM for each timespan in the entire period, and each CIM should be under the same framework which we constructed in the previous step.
    \item Third, we apply consistency algorithm for each CIM and acquire consistent scenarios in each timespan. Each consistent scenarios have a specific scenario weight.
    \item Last, for each timespan, we choose the scenario that have the highest scenario weight and connect these scenarios as a chain. This chain can depict the most probable evolve path of the system.
\end{enumerate}

\section{Discussion}
The workflow we raised in section 2 provides a new approach to deal with CIB in multiple timespans. In this section, we will discuss the details and the limitations of this approach.

In the first step, the process is as same as the original CIB, and the researchers can determine the system and timespans by organizing expert consultation. This descriptors-states framework will be applied in every timespan in the second period. This step can combine with other extension methods on choosing of descriptors and states, such as organizing stakeholder workshops\cite{kurniawan2022}.

The second step is the most onerous part of the entire workflow since it requires the researchers to construct all the CIMs altogether. However, the advantage of this approach is also obvious, as it decreases the number of expert consultation hold, and all the researchers and experts could focus on the analyze of interrelationships of each pair of descriptors in different timespans. The fix of all the judgement sections faces the same problem as the dynamic CIB, which requires researchers to investigate the future relationships between the descriptors. But by combining with literature review and statistic analyze, the previous investigation is a feasible operation, and it is easier than directly find the future scenarios. Therefore, since the interrelationship of each pair of descriptors may be different in each timespan, a single judgement section or judgement cell can be empty in a timespan, but non-empty in another.

For a specific CIM, if we use the consistency algorithm to check all the scenarios, we will only get consistent scenarios. However, since we need the ‘scenario weight’ of each consistent scenario for the construction of scenario chain in the last step, we need a strategic called ‘succession’, which the detail can be found in Weimer-Jehle (2009)\cite{weimer2009}. There are four main succession rules to reach a consistent scenario or cyclic attractor, so here we want to highlight that, first, we do not need cyclic attractors in our workflow as our goal is to construct a fixed scenario chain. Second, the four succession rules are global, incremental, local, and adiabatic, which can correspond to four types of evolution policies:
\begin{itemize}
    \item Global succession: flooding and radical policy, trying to turn all inconsistent descriptors into consistent at once.
    \item Incremental succession: milder than global succession, trying to solve all the inconsistency step by step.
    \item Local succession: milder policy, trying to solve the problem one by one.
    \item Adiabatic succession: similar to local succession, but make a list before each succession and solve the first problem every time.
\end{itemize}
Third, choosing different succession rules may acquire different scenario weight, thus requires the policy makers to choose the ‘policy type’ before the last step.

In the workflow of section 2, the last step simply uses the scenario weight to construct the scenario chain, and this is the basic approach. However, in practical using, the basic approach may encounter some problem. Try to think a descriptors-states system and corresponding CIM, which all the states of the descriptors and all the judgement sections are symmetric. Thus, for extreme good and extreme bad consistent scenarios of this CIM, the scenario weight of them should be the same, then the problem emerges. When this property extends to multiple timespans, our scenario chain may contain extreme good and extreme bad scenarios in adjacent timespans.

Therefore, we may need to overlay a filter or a kind of manually assigned weight for different research purposes. For the research of development or recession pathways, we can assign a specific value for each state of every descriptor. We can assign larger value for better states in the research of development pathways and vice versa. After the step 3, we plus all the assigned value for each consistent scenarios as a manual weight, then multiply by scenario weight as a compounded weight before apply step 4. Another approach is to organize another expert consultation and vote to the consistent scenarios to create a filter or a weight. In conclusion, step 4 is the most flexible step of our workflow, which depends on the research topic.

\section{Summary}
In this article, we raised another possible approach to deal with CIB method in multiple timespans. The concise workflow and flexible approach can give policy makers more convenience during the research. However, the flexibility in the step 4 may still require us to find a more fixed way to weight the consistent scenarios in further research to construct scenario chain.

\section*{Appendix Formalization of scenarios aggregation in multi-level CIB}
In this part, we will formalize the algorithm of scenarios aggregation in multi-level CIB\cite{vogele2017}.

Given a descriptor set $\Delta=\{D_1, D_2, \cdots, D_N\}, N\in \mathbb{N}^+$ and the states $D_k=\{D_k^1,D_k^2,\cdots,D_k^{s_k}\},s_k\in\mathbb{N_+}$ of every descriptors, $s_k$ denotes the number of the states of a descriptor, we call the descriptor set and the states of them a \textbf{system}. Denote the \textbf{direct impact} of $D_i^{s_i}$ to $D_j^{s_j}$ as $d_{ij}(s_i,s_j)$, we can denote CIM as a block matrix $(D_{ij})$, call $D_{ij}$ as the \textbf{judgement cell} of $D_i$ to $D_j$. We have
\begin{equation*}
    D=\begin{bmatrix}
    O & D_{12} & \cdots & D_{1N}\\
    D_{21} & O & \cdots & D_{2N}\\
    \vdots & \vdots & \ddots & \vdots\\
    D_{N1} & D_{N2} & \cdots & O
    \end{bmatrix}
    ,\;{\text {which}}\;D_{ij}=\begin{bmatrix}
    d_{ij}(1,1) & d_{ij}(1,2) & \cdots & d_{ij}(1,s_j)\\
    d_{ij}(2,1) & d_{ij}(2,2) & \cdots & d_{ij}(2,s_j)\\
    \vdots & \vdots & \ddots & \vdots\\
    d_{ij}(s_i,1) & d_{ij}(s_i,2) & \cdots & d_{ij}(s_i,s_j)
    \end{bmatrix}
    ,i\neq j.
\end{equation*}

Given a descriptor set $\Delta=\{D_1, D_2, \cdots, D_N\}, N\in \mathbb{N}^+$, consider the finite splitting of $\Delta$ as $\Bar{\Delta}=\{\Delta_i\}^X_{i=1}$, such that $\Delta=\bigcup^X_{i=1}\Delta_i$, and for each $\Delta_i\in\Bar{\Delta}$, there exists a $\Delta_j\in\Bar{\Delta}$, such that $\Delta_i\cap\Delta_j\neq\varnothing$. We call $\Delta_i$ as the \textbf{sub-descriptor set} of $\Delta$, moreover, the sub-descriptor set and the states of them called a \textbf{subsystem} corresponding to the sub-descriptor set. Denote $T=\{D_t|D_t\in\Delta_i\cap\Delta_j,i,j=1,2,\cdots,X\}$ as \textbf{transitional descriptor set}. Use $|\cdot|$ to denote the number of elements of a set. We suppose that for all $\Delta_i\cap\Delta_j=\varnothing$, for each $D_i\in\Delta_i,D_j\in\Delta_j$, we have $D_{ij}=O$, it means the descriptors in non-overlapped sub-descriptor set do not have direct impact with each other.

As we do not change the state of the descriptors when split the system into subsystems, we call the CIM corresponding to the subsystem corresponding to the sub-descriptor set as the CIM corresponding to the sub-descriptor set for short.

For all the sub-descriptor set split from the descriptor set, we construct all the CIM corresponding to the sub-descriptor set, then, by applying the consistency algorithm, we acquire each \textbf{consistent scenario set} $S_i$. Denote $\Xi=\{\{z_1, z_2, \cdots, z_X\}|z_i\in S_i,i=1,2,\cdots X\}$ as the \textbf{combinatorial consistent scenario set} between the sub-descriptor sets, which $\xi=\{z_1, z_2, \cdots, z_X\}$ called a \textbf{combinatorial consistent scenario} between the sub-descriptor sets, we call it as a \textbf{combinatorial} for short.

When constructing the CIM of $\Delta_i$, if it exists $D_i,D_j\in T$ and $\Delta_m,\Delta_n\in\Delta$ such that it holds $D_i,D_j\in\Delta_m$ and $D_i,D_j\in\Delta_n$ at the same time, then we must ensure that the judgement sections of $D_{ij}$ in the CIM of $\Delta_m,\Delta_n$ must be the same.

For $T$ and $\Xi$, if $\xi=\{z_1, z_2, \cdots, z_X\}\in\Xi$ meets
\begin{equation*}
    \bigcup_{1\leq i,j\leq X,i\neq j}({z_i}\cap{z_j})=|T|,\;{\text {which}}\;z_i\in \xi
\end{equation*}
then we say that combinatorial $\xi$ can be \textbf{aggregated} into a consistent scenario in the scenario space of $\Delta$, and for all such combinatorial $\xi=\{z_1, z_2, \cdots, z_X\}\in\Xi$, the consistent scenarios of $\Delta$ is $z=\bigcup_{i=1}^X z_i$.

\bibliography{timecib}


\end{document}